\documentclass[aps,prb,reprint,superscriptaddress]{revtex4-1}
\usepackage{amsmath}
\usepackage{graphicx}
\usepackage{siunitx}
\usepackage{hyperref}

\graphicspath{{Figures/}}
\hypersetup{
	colorlinks=true,
	linkcolor=blue,
	filecolor=black,      
	urlcolor=blue,
	citecolor=blue,
}

\newcommand{\SrIrO}{Sr\textsubscript{2}IrO\textsubscript{4}}
\newcommand{\SrLaIrO}{(Sr\textsubscript{1-x}La\textsubscript{x})\textsubscript{2}IrO\textsubscript{4}}
\newcommand{\LaCuO}{La\textsubscript{2}CuO\textsubscript{4}}

\begin{document}

\title{Momentum-resolved lattice dynamics of parent and electron-doped \SrIrO{}} 

\author{C.~D.~Dashwood}
\email{cameron.dashwood.17@ucl.ac.uk}
\affiliation{London Centre for Nanotechnology and Department of Physics and Astronomy, University College London, London WC1E 6BT, UK}

\author{H.~Miao}
\affiliation{Department of Condensed Matter Physics and Materials Science, Brookhaven National Laboratory, Upton, New York 11973, USA}

\author{J.~G.~Vale}
\affiliation{London Centre for Nanotechnology and Department of Physics and Astronomy, University College London, London WC1E 6BT, UK}

\author{D.~Ishikawa}
\affiliation{Materials Dynamics Laboratory, RIKEN SPring-8 Center, RIKEN, Sayo Hyogo 697-5148, Japan}

\author{D.~A.~Prishchenko}
\affiliation{Department of Theoretical Physics and Applied Mathematics, Ural Federal University, 19 Mira Street, Ekaterinburg 620002, Russia}

\author{V.~V.~Mazurenko}
\affiliation{Department of Theoretical Physics and Applied Mathematics, Ural Federal University, 19 Mira Street, Ekaterinburg 620002, Russia}

\author{V.~G.~Mazurenko}
\affiliation{Department of Theoretical Physics and Applied Mathematics, Ural Federal University, 19 Mira Street, Ekaterinburg 620002, Russia}

\author{R.~S.~Perry}
\affiliation{London Centre for Nanotechnology and Department of Physics and Astronomy, University College London, London WC1E 6BT, UK}

\author{G.~Cao}
\affiliation{Department of Physics, University of Colorado at Boulder, Boulder, Colorado 80309, USA}

\author{A.~de la Torre}
\affiliation{Institute for Quantum Information and Matter and Department of Physics, California Institute of Technology, Pasadena, California 91125, USA}
\affiliation{Department of Quantum Matter Physics, University of Geneva, 24 Quai Ernest-Ansermet, 1211 Geneva 4, Switzerland}

\author{F.~Baumberger}
\affiliation{Department of Quantum Matter Physics, University of Geneva, 24 Quai Ernest-Ansermet, 1211 Geneva 4, Switzerland}

\author{A.~Q.~R.~Baron}
\affiliation{Materials Dynamics Laboratory, RIKEN SPring-8 Center, RIKEN, Sayo Hyogo 697-5148, Japan}

\author{D.~F.~McMorrow}
\address{London Centre for Nanotechnology and Department of Physics and Astronomy, University College London, London WC1E 6BT, UK}

\author{M.~P.~M.~Dean}
\email{mdean@bnl.gov}
\affiliation{Department of Condensed Matter Physics and Materials Science, Brookhaven National Laboratory, Upton, New York 11973, USA}

\date{\today}

\begin{abstract}
The mixing of orbital and spin character in the wave functions of the $5d$ iridates has led to predictions of strong couplings among their lattice, electronic and magnetic degrees of freedom. As well as realizing a novel spin-orbit assisted Mott-insulating ground state, the perovskite iridate \SrIrO{} has strong similarities with the cuprate \LaCuO{}, which on doping hosts a charge-density wave that appears intimately connected to high-temperature superconductivity. These phenomena can be sensitively probed through momentum-resolved measurements of the lattice dynamics, made possible by \SI{}{\milli\electronvolt}-resolution inelastic x-ray scattering. Here we report the first such measurements for both parent and electron-doped \SrIrO{}. We find that the low-energy phonon dispersions and intensities in both compounds are well described by the same nonmagnetic density functional theory calculation. In the parent compound, no changes of the phonons on magnetic ordering are discernible within the experimental resolution, and in the doped compound no anomalies are apparent due to charge-density waves. These measurements extend our knowledge of the lattice properties of \SrLaIrO{} and constrain the couplings of the phonons to magnetic and charge order.
\end{abstract}

\maketitle

\section{Introduction}

The delicate balance of spin-orbit coupling (SOC), crystal fields and electron correlations (U) in the $5d$ iridates makes them a fruitful class of materials in the search for novel electronic and magnetic phases \cite{Jackeli2009, Witczak-Krempa2014, Cao2018, Bertinshaw2018}. Most prominently, the layered perovskite \SrIrO{} has been shown to be a spin-orbit Mott insulator where the orbital degeneracy of the Ir\textsuperscript{4+} $t_{2g}$ levels is lifted by SOC, enabling a moderate $U \sim$ \SI{2}{\electronvolt} to open a charge gap \cite{Kim2008, Kim2009}. Moreover, it has striking structural, electronic and magnetic similarities to the parent of the cuprate high-temperature superconductors \LaCuO{} \cite{Crawford1994, Wang2011, Kim2012, Boseggia2013}. Doping the bulk of \SrIrO{} with electrons leads to the suppression of long-range antiferromagnetic order \cite{Pincini2017}, while surface-doping has been shown to produce Fermi arcs \cite{Kim2014} and a low-temperature gap with \textit{d}-wave symmetry \cite{Kim2016}.

One difference from the cuprates is the orbital character of the $j_\textrm{eff} = 1/2$ wave function, which results in the couplings between the pseudospins being highly sensitive to lattice geometry \cite{Jackeli2009}. Recent theoretical \cite{Liu2019} and experimental \cite{Porras2019} works have shown that this coupling is crucial to understanding the magnetic structure and in-plane magnon gap of \SrIrO{}. Lattice distortions can result in significant admixture of the $j_\textrm{eff} = 3/2$ wave function into the $j_\textrm{eff} = 1/2$ ground state, which has led to expectations of strong interactions among lattice, orbital, and magnetic excitations in the iridates.

Changes in the frequencies and linewidths of phonon modes on magnetic ordering are seen in a variety of $5d$ transition metal oxides, including the pyrochlore iridate Y\textsubscript{2}Ir\textsubscript{2}O\textsubscript{7} \cite{Son2019}, and the osmates NaOsO\textsubscript{3} \cite{Calder2015} and Ca\textsubscript{2}Os\textsubscript{2}O\textsubscript{7} \cite{Sohn2017}. The frequency shift in NaOsO\textsubscript{3} is the largest measured in any material, at \SI{5}{\milli\electronvolt} \cite{Calder2015}. Gretarsson \textit{et al.} \cite{Gretarsson2016} conducted Raman measurements on \SrIrO{}, finding a broadening of $\sim$\SI{1}{\milli\electronvolt} and Fano asymmetry in the $A_{1g}$ phonon mode at the zone center above $T_N \approx$ \SI{240}{\kelvin}, that is indicative of coupling between the lattice and a continuum of pseudospin fluctuations due to unquenched orbital dynamics.

A notable feature of the underdoped cuprates is the appearance of charge-density wave (CDW) order above a critical doping, which appears to be connected to the superconductivity in these compounds \cite{Tranquada1995, Comin2016}. CDW order has long been proposed to be an intrinsic instability of doped Mott insulators \cite{Zaanen1989, Poilblanc1989}, but despite reports of spin-density wave (SDW) order \cite{Chen2018} in doped \SrIrO{} and a dynamic CDW-like instability in its bilayer cousin Sr\textsubscript{3}Ir\textsubscript{2}O\textsubscript{7} \cite{Chu2017, Jin2019}, as yet there has been no evidence for a CDW in doped \SrIrO{}. The presence of CDW order can be inferred, \textit{inter alia}, from the softening of phonon modes around the CDW wave vector \cite{Weber2011, Reznik2006, Miao2018}.

\begin{figure*}
	\centering
	\includegraphics[width=\linewidth]{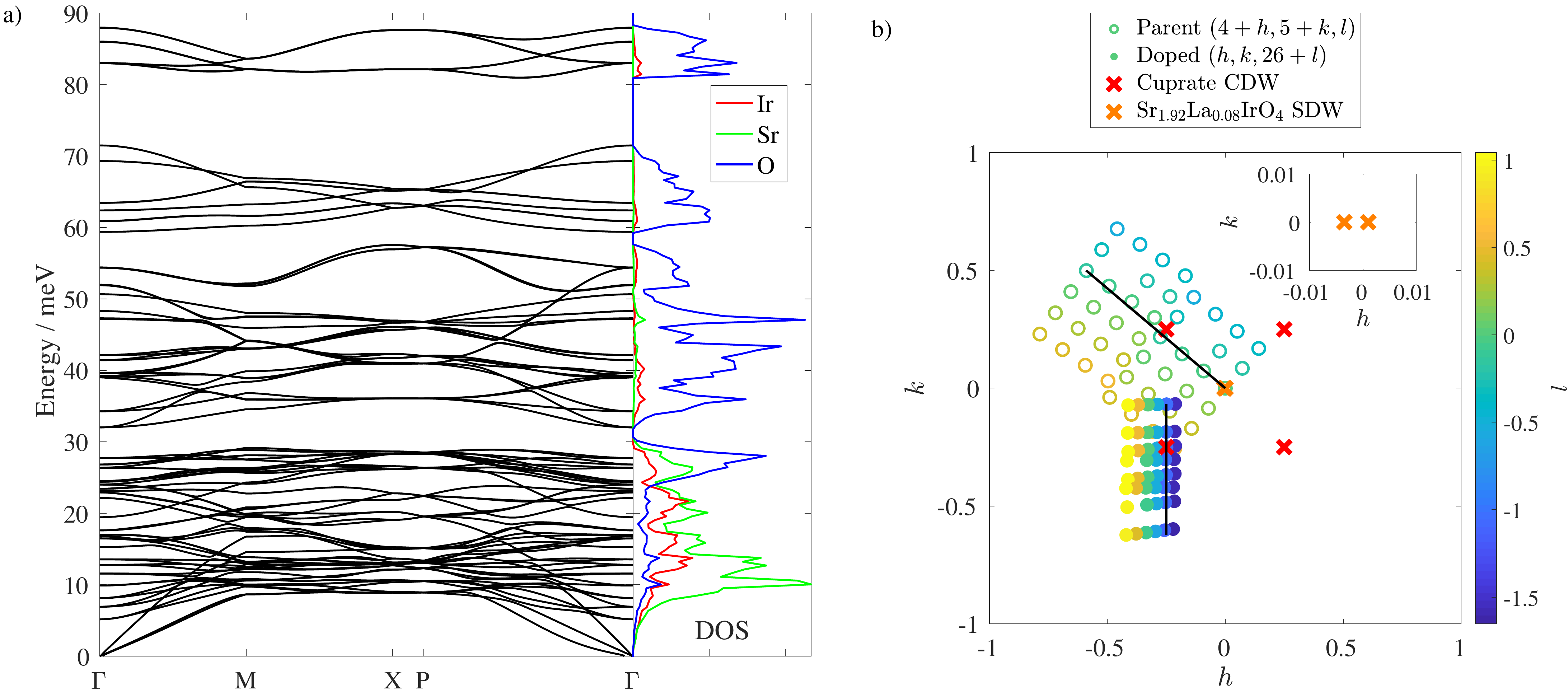}
	\caption{(a) Phonon band structure (black lines) and projected phonon DOS for Ir (red), Sr (green) and O (blue), calculated in the LDA on a $2 \times 2 \times 1$ supercell. The high-symmetry points of the $I4_1/acd$ space group are defined as $\Gamma = (0, 0, 0)$, $M = (0.5, 0, 0)$, $X = (0.5, 0.5, 0)$, and $P = (0.5, 0.5, 0.5)$. (b) Brillouin zone map (structural and magnetic zones are equivalent in the $I4_1/acd$ notation) showing the measured $\mathbf{Q}$ points for the parent compound relative to $(4, 5, 0)$ as empty circles, and for the doped compound relative to $(0, 0, 26)$ as filled circles. The points have been projected onto the $h - k$ plane, with $l$ values indicated by color. The red crosses show the equivalent in-plane wave vector of the cuprate CDW \cite{Miao2018} and orange crosses those of the purported SDW in doped \SrIrO{} \cite{Chen2018} (shown magnified in the inset). The black lines indicate the reciprocal space directions for which data are shown in this manuscript, with spectra at the other $\mathbf{Q}$ points contained in the supplemental material \cite{supplemental}.}
	\label{fig:DFT}
\end{figure*}

There is therefore a clear interest in momentum-resolved measurements of the phonons in \SrIrO{}, extending the previous zone-center studies. Nonresonant inelastic x-ray scattering (IXS) allows such measurements to be performed with $\sim$\SI{1}{\milli\electronvolt} energy resolution and $\sim$0.01 reciprocal lattice units (r.l.u.) momentum resolution across a large volume of reciprocal space \cite{Baron2009, Baron2016}. We have performed extensive IXS measurements on parent and La-doped \SrIrO{} across regions of reciprocal space carefully chosen to maximize the signatures of coupling to magnetic or CDW order. In the parent compound, we find that our IXS spectra are well reproduced by a nonmagnetic density functional theory (DFT) calculation, which allows us to identify the dominant atomic displacements and interrogate modes with strong modulation of the magnetic exchange pathways. Evaluation of the dynamic structure factor from this DFT calculation allows us to quantify the expected temperature dependence of the spectra and reveal that there are no changes due to magnetic ordering within our experimental resolution. We observe minimal changes to the phonons on doping, with the dispersions well reproduced by the same DFT calculation. A careful fitting of the IXS spectra reveals no anomalies due to CDW order between \SI{250}{\kelvin} and \SI{9}{\kelvin}.

\section{Methods}

Single crystals of both parent ($x = 0$) and 5\% doped ($x \approx 0.05$) \SrLaIrO{} were flux grown using standard methods and characterized by energy-dispersive x-ray spectroscopy, resistivity, and susceptibility measurements \cite{delaTorre2015}. The crystalline quality of the samples was checked during the IXS measurements, with mosaics of around \SI{0.02}{\degree} for the parent compound and \SI{0.05}{\degree} for the doped. Throughout this manuscript, we use the $I4_1/acd$ space group with $a = b =$ \SI{5.50}{\angstrom} and $c =$ \SI{25.79}{\angstrom}. The true space group of \SrIrO{} is now known to be $I4_1/a$ \cite{Ye2013, Dhital2013, Torchinsky2015, Ye2015}, although this makes negligible difference to the phonon dispersions (see supplemental material \cite{supplemental}).

DFT calculations were performed using the plane-wave basis projector augmented wave method \cite{Blochl1994} as implemented in the Vienna \textit{ab-initio} Simulation Package (\textsc{vasp}) \cite{Kresse1996}. The exchange-correlation functional was treated in the local density approximation (LDA) \cite{Perdew1992}, with unit cell relaxations carried out over an $8 \times 8 \times 2$ reciprocal lattice mesh. The force constants were calculated over a $4 \times 4 \times 2$ mesh using a $2 \times 2 \times 1$ supercell. The phonon frequencies and eigenvectors were then obtained with the \textsc{phonopy} package \cite{Togo2015} using an $11 \times 11 \times 11$ mesh for the Debye-Waller factor, and these were used to calculate the dynamic structure factor $S(\mathbf{Q}, \omega)$ (see supplemental material \cite{supplemental}). The resulting phonon band structure and projected density of states (DOS) is shown in Fig.~\ref{fig:DFT}(a). As expected, the modes involving motion of mostly the heavier Sr and Ir atoms lie at lower energies, while the modes involving lighter O atoms dominate above \SI{30}{\milli\electronvolt}. The calculated energies at the $\Gamma$ point compare well to previous Raman and infrared spectroscopy studies \cite{Samanta2018, Gretarsson2017, Propper2016} (see supplemental material \cite{supplemental}).

Calculations were also performed including the effects of SOC+U with the full noncollinear magnetic structure \cite{Ye2013}, but the computational complexity of this meant that the supercell size had to be reduced to $1\times1\times1$, at which point agreement between the calculated $S(\mathbf{Q}, \omega)$ and IXS spectra away from the zone center became unsatisfactory (see Fig. \ref{fig:parent_spectra} for a comparison with the above LDA calculation on a $2\times2\times1$ supercell, and the Appendix for further discussion).

High-resolution IXS measurements of the phonon dispersions were performed at beamline BL43LXU of the SPring-8 synchrotron in Japan \cite{Baron2010}. The incident energy was set to \SI{21.75}{\kilo\electronvolt} and the $(11, 11, 11)$ reflection of Si was used as both a monochromator and analyzer, giving an energy resolution of around \SI{1.5}{\milli\electronvolt} (depending on analyzer). A $4 \times 6$ analyzer array allowed the simultaneous measurement of many momentum transfers, so that a large area of the Brillouin zone, shown in Fig.~\ref{fig:DFT}(b), could be surveyed despite the long counting times necessitated by the high energy and momentum resolutions. Given these counting times, we chose to investigate the high-intensity phonon modes below \SI{30}{\milli\electronvolt}, giving us the best chance of observing the small changes expected from magnetic or CDW order.

The parent compound was measured in transmission with the $[1, 1, 0]$ and $[0, 0, 1]$ reciprocal directions in the scattering plane, allowing access to in-plane momentum transfers in order to maximize the intensity of low-energy modes that involve modulation of the Ir-O-Ir superexchange bond. The vertical columns of the analyzer array traced out adjacent trajectories along $[\bar{1}, 1, 0]$ out from the $(4, 5, 0)$ magnetic Bragg peak position [open circles in Fig.~\ref{fig:DFT}(b)]. As well as avoiding points near the structural Bragg peaks $(4, 4, 0)$ and $(4, 6, 0)$ at which the IXS spectra would be dominated by strong elastic contributions (which does not occur at the magnetic Bragg peak off resonance), this is also the direction along which the dynamic CDW is expected in Sr\textsubscript{3}Ir\textsubscript{2}O\textsubscript{7} \cite{Jin2019}. The atomic displacements of modes with significant IXS intensity were calculated using DFT for a range of other $\mathbf{Q}$ points in the $a - b$ plane, but no modes could be found with significantly larger modulation of the Ir-O-Ir bond that we would expect to be more strongly influenced by magnetism. The black lines in Fig.~\ref{fig:DFT}(b) indicate the points with equal $l$ for which spectra are shown in this manuscript. Spectra for the other points are contained in the supplemental material \cite{supplemental}.

The strongest coupling to CDW order is usually found for phonon modes whose displacements mirror those of the CDW, and for this reason the early work investigating such coupling in the cuprates focused on in-plane modes with strong distortion of the Cu-O bond \cite{McQueeney1999, Uchiyama2004, Reznik2006, Reznik2008, Graf2008}. These modes are at high energies, however, with IXS intensities too low to allow practical measurement. We therefore focused on low-energy modes involving motion of all atoms in the unit cell, which should also show appreciable coupling. This was confirmed by recent IXS measurements on La\textsubscript{1.875}Ba\textsubscript{0.125}CuO\textsubscript{4}, which found that the coupling is strongest for low-energy modes with the large $c$-axis displacements \cite{Miao2018}. These modes are enhanced by having a large out-of-plane momentum transfer ($l$), so for the doped iridate sample we used a reflection geometry with the $[1, 0, 0]$ and $[0, 0, 1]$ directions in the scattering plane, measuring $\mathbf{Q}$ points in the $(0, 0, 26)$ Brillouin zone [filled circles in Fig.~\ref{fig:DFT}(b)]. A vertical column of the analyzer array then followed the $[0, \bar{1}, 0]$ direction (black line) through the equivalent cuprate CDW in-plane wave vector $(-0.25, -0.25)$ (filled red circles).

A consequence of the analyzer geometry is that $l$ varies horizontally across the array, as indicated by the color of the points in Fig.~\ref{fig:DFT}(b). The layered nature of \SrIrO{}, however, means that the electronic \cite{Wilkins2011} and magnetic \cite{Kim2012} behavior of interest is only weakly $l$ dependent. We therefore set the analyzer slits to \SI{40}{}$\times$\SI{80}{\milli\meter} to improve the in-plane momentum resolutions while relaxing the out-of-plane resolution. The momentum resolutions are reported below for each set of measurements.

Other $\mathbf{Q}$ points where one might expect the presence of phonon anomalies is at the intersections of the phonon and magnon dispersions. Unfortunately, the high spin-wave velocity and slight gap \cite{Pincini2017} places it at energies above those of the phonon modes measured in this work.

\section{Results}

\subsection{Parent compound}

\begin{figure}
	\centering
	\includegraphics[width=\linewidth]{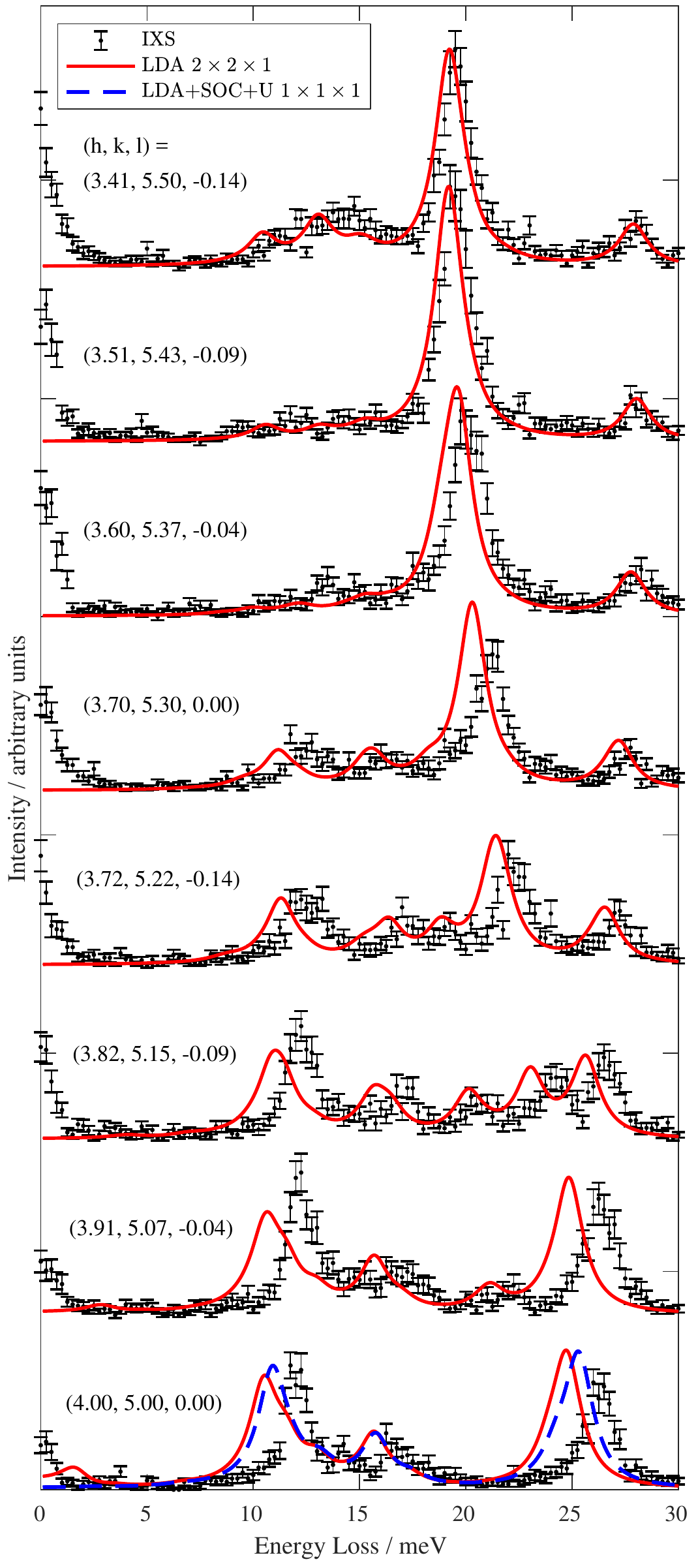}
	\caption{Representative IXS spectra at momenta along $[\bar{1}, 1, 0]$ out from the magnetic Bragg peak position at $(4, 5, 0)$ in the parent compound at \SI{100}{\kelvin} (black points) compared to $S(\mathbf{Q}, \omega)$ calculated with DFT in the LDA on $2 \times 2 \times 1$ supercell (red solid lines) and LDA+SOC+U on a $1 \times 1 \times 1$ supercell at the zone center (blue dashed line). While the LDA calculation matches the measured intensities well across the Brillouin zone, it underestimates the mode energies toward the zone center. The spectra are offset vertically for clarity.}
	\label{fig:parent_spectra}
\end{figure}

Figure \ref{fig:parent_spectra} shows a series of representative IXS spectra along the $[\bar{1}, 1, 0]$ direction out from the magnetic Bragg peak position $(4, 5, 0)$ in the parent compound at 100 K. The average momentum resolutions along each direction are $\boldsymbol{\delta} \mathbf{Q} = (0.06, 0.07, 0.13)$~r.l.u. The IXS spectra (black points) are overlaid with $S(\mathbf{Q}, \omega)$ calculated with DFT in the LDA on a $2 \times 2 \times 1$ supercell (red lines). At all momenta, the calculation reproduces the relative intensities of the modes reasonably well, with the consistent underestimate of the mode frequencies improving further out into the Brillouin zone.

The calculation allows us to identify the atomic motion associated with each prominent peak of the IXS spectra. As expected from the in-plane momentum transfer, the modes mostly involve atomic motions in the $a-b$ plane. At the zone center, the prominent mode with an LDA energy of \SI{10.5}{\milli\electronvolt} has large displacements of the Sr atoms with smaller Ir and O motion, while the mode at \SI{24.7}{\milli\electronvolt} has dominant in-plane O motion along with smaller out-of-plane Sr oscillations. Toward the zone boundary, the mode around \SI{19}{\milli\electronvolt} has roughly equal in-plane displacements of all atoms. Animations of these modes can be found in the supplemental material \cite{supplemental}.

\begin{figure}
	\centering
	\includegraphics[width=\linewidth]{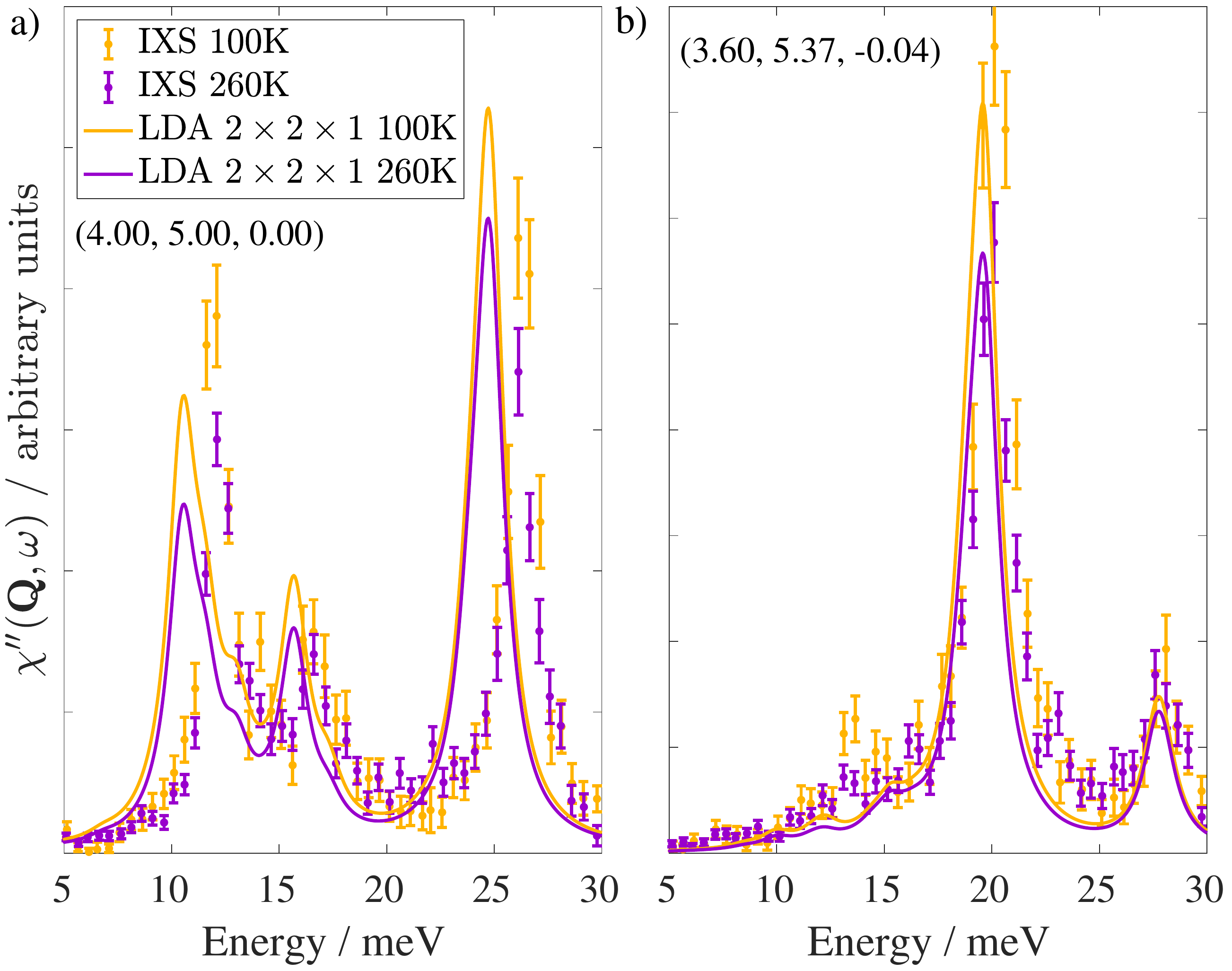}
	\caption{Bose-factor corrected IXS spectra in the parent compound (points) and $\chi''(\mathbf{Q}, \omega)$ calculated with DFT in the LDA on $2 \times 2 \times 1$ supercell (solid lines) at \SI{100}{\kelvin} (orange) and \SI{260}{\kelvin} (purple) for momentum transfers of (a) $(4, 5, 0)$ and (b) $(3.60, 5.37, -0.04)$. There is no apparent softening or changes in linewidth for any of the modes, while the intensity changes are fully accounted for by the Debye-Waller factor.}
	\label{fig:parent_temp_comp}
\end{figure}

While there is unlikely to be any detectable influence from magnetism on the low-energy mode with dominant Sr motion, the higher-energy zone-center mode involves significant changes to the angle of the Ir-O-Ir bond that is responsible for magnetic superexchange. There are no apparent discrepancies between the experimental data and calculated spectrum for the high-energy mode that are not also present for the low-energy mode, however.

As mentioned above, we also performed DFT calculations including SOC+U with the full magnetic structure of \SrIrO{}. At the zone center, where the reduction to a $1\times1\times1$ supercell size should make a minimal difference (see the Appendix for further discussion), this simply causes a $\sim$\SI{0.5}{\milli\electronvolt} increase in the predicted energies of both modes with very little change to the intensities (blue dashed line in Fig.~\ref{fig:parent_spectra}).

We repeated these measurements at \SI{260}{\kelvin}, above $T_N \approx$ \SI{240}{\kelvin}, in order to search for any changes caused by long-range magnetic ordering. To reliably compare spectra at different temperatures, the imaginary part of the dynamic susceptibility was calculated by subtracting the elastic line and dividing through by the Bose factor \cite{Miao2018}
\begin{equation}
	\displaystyle \chi''(\mathbf{Q}, \omega) = S(\mathbf{Q}, \omega) \left(1 - e^{-\hbar \omega / (k_BT)}\right).
\label{equ:Bose_correction}
\end{equation}
Figure \ref{fig:parent_temp_comp} shows $\chi''$ at \SI{100}{\kelvin} and \SI{260}{\kelvin} for two different representative momentum transfers. The intensity of the modes in the \SI{260}{\kelvin} spectra are lower than those at \SI{100}{\kelvin} due to the reduced Debye-Waller factor, as can be seen through comparison with $\chi''(\mathbf{Q}, \omega)$ calculated with DFT. The observed hardening of the phonon modes of $\sim$\SI{0.5}{\milli\electronvolt} can be attributed to reduced anharmonic phonon-phonon interactions on cooling. There is no clear evidence for changes in the linewidth through $T_N$, as seen in Raman measurements by Gretarsson \textit{et al.} \cite{Gretarsson2016} which would be indicative of coupling to spin fluctuations. Although the asymmetric broadening of the $A_{1g}$ mode at \SI{23}{\milli\electronvolt} reported by Gretarsson \textit{et al.} \cite{Gretarsson2016} is of a magnitude comparable to our energy resolution, and would therefore be visible in our data, it should be noted that this particular mode has vanishing IXS intensity at the $\mathbf{Q}$ points measured here.

The spectra and temperature comparisons for the other momentum transfers shown in Fig.~\ref{fig:DFT}(c) can be found in the supplemental material \cite{supplemental}. The same conclusions discussed above can be made for all of these data sets.

\subsection{Doped compound}

In the cuprates, a signature of the presence of CDW order is the softening and linewidth changes of phonon peaks in IXS spectra at the CDW wave vector. IXS measurements on (La\textsubscript{1-x}Ba\textsubscript{x})\textsubscript{2}CuO\textsubscript{4} with $x =$ 0.048 - 0.063 revealed that precursor CDW fluctuations are responsible for a broadening and softening of the phonon modes \cite{Pintschovius1991, McQueeney1999, Uchiyama2004, Reznik2006, Reznik2008, Graf2008, dAstuto2008, Miao2018}. On the onset of CDW ordering the softening is still present while there is a sharp reduction in the phonon linewidths.

\begin{figure}
	\centering
	\includegraphics[width=\linewidth]{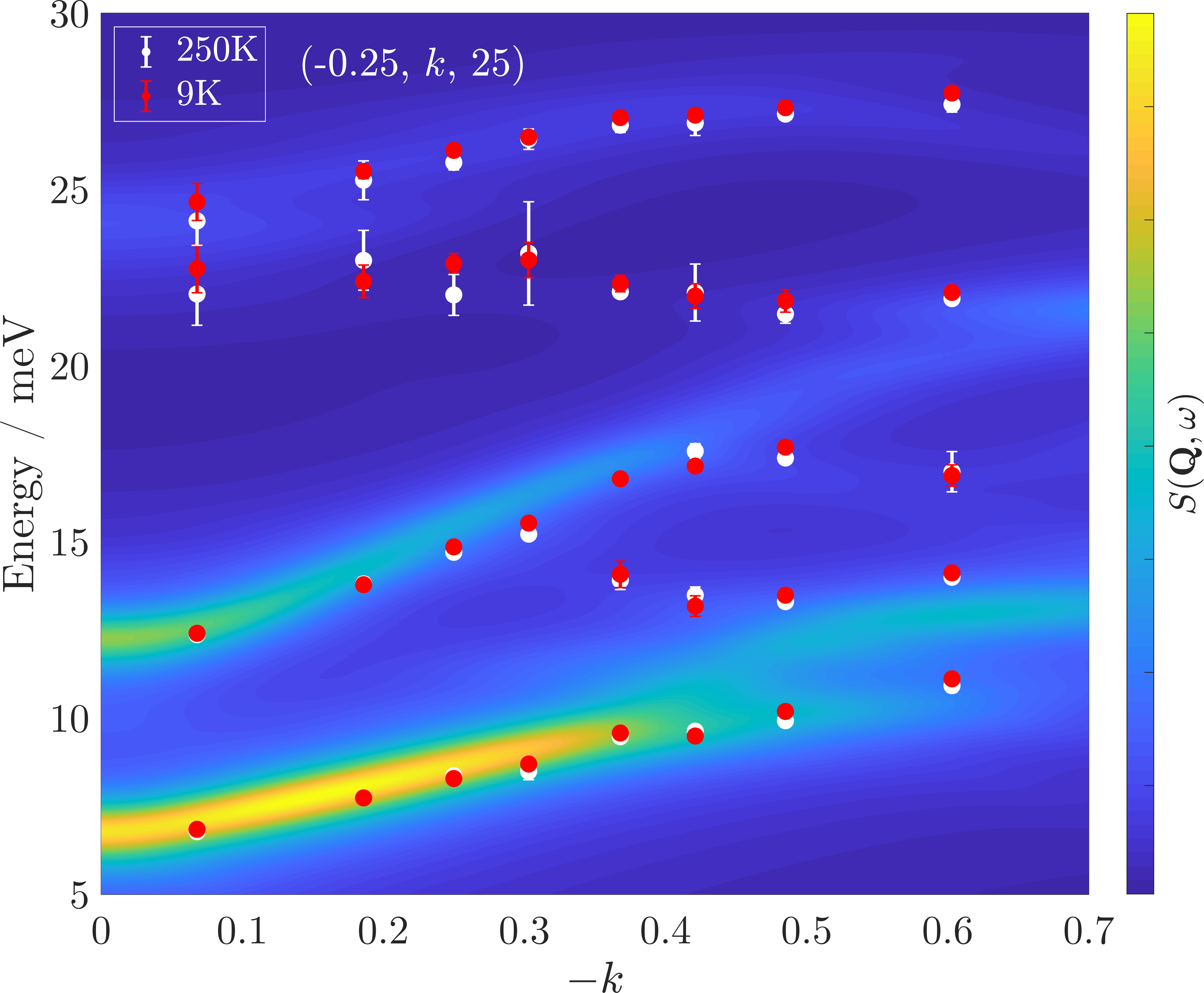}
	\caption{Phonon dispersions at \SI{250}{\kelvin} (white circles) and \SI{9}{\kelvin} (red circles) extracted from fits to the IXS spectra of the doped compound. These are overlaid on a colormap of the dynamic structure factor calculated with DFT in the LDA on $2 \times 2 \times 1$ supercell. Vertical error bars represent the statistical errors from fitting. The dispersions are identical within error at the two temperatures, with no apparent softening, and agree well with the calculation.}
	\label{fig:doped_dispersion}
\end{figure}

To investigate whether CDW order is present in electron-doped \SrIrO{} in analogy with the hole-doped cuprates, we performed IXS measurements on 5\% La-doped \SrIrO{} along the $[0, \bar{1}, 0]$ direction through $(-0.25, -0.25, 25)$ with average momentum resolutions of $(0.04, 0.04, 0.36)$~r.l.u. In order to extract the phonon dispersions, the IXS spectra were fitted to a sum of damped harmonic oscillator line shapes $\chi_j''$ weighted by the Bose factor and convoluted with a Voigt resolution function $R$
\begin{equation}
	\displaystyle S(\mathbf{Q}, \omega) = \sum_{j} \frac{\chi_j''(\mathbf{Q}, \omega)}{1 - e^{-\hbar \omega / (k_BT)}} * R(\omega)
\label{equ:fitting}
\end{equation}
plus an additional Voigt function for the quasielastic peak (see supplemental material \cite{supplemental} for a representative fit). The fitted dispersions at \SI{250}{\kelvin} and \SI{9}{\kelvin} are shown in Fig.~\ref{fig:doped_dispersion} as white and red circles respectively, overlaid on a colormap of $S(\mathbf{Q}, \omega)$ from the same LDA calculation on a $2 \times 2 \times 1$ supercell as above.

This nonmagnetic DFT calculation actually provides a better description of the metallic ground state of the doped sample, in which long-range magnetic order is destroyed by the free carriers \cite{Pincini2017}, and so as expected there is good agreement between the fitted and calculated dispersions. As for the parent compound, the calculations allow us to identify the atomic motions associated with each mode, which at these wave vectors involve significant out-of-plane displacements for all three elements.

Crucially, the fitted dispersions are identical (within one standard deviation) at both temperatures, with no anomalies present at the equivalent in-plane wave vector $(-0.25, -0.25)$ to the cuprate CDW. Anomalies are also not apparent at any other $\mathbf{Q}$ points measured in this work (see supplemental material \cite{supplemental}). It should be noted, however, that this does not preclude the presence of a CDW for other doping levels (the purported spin density wave only occurs over a very narrow doping range \cite{Chen2018}), at wave vectors away from those measured here, or one that couples to a phonon modes with low IXS intensity.

A further signature of CDW order in the cuprates is visible in the intensity of the quasielastic peak centered at zero energy in the IXS spectra. Le Tacon \textit{et al.} \cite{leTacon2014} saw a contribution to the integrated intensity of this peak in the underdoped cuprate YBa\textsubscript{2}Cu\textsubscript{3}O\textsubscript{6.6} over a narrow momentum range around the CDW wave vector and over a broad temperature range around the CDW transition temperature. At both 9 K and 250 K, however, the fitted integrated intensity of the quasielastic peak in our IXS spectra varies smoothly with $\mathbf{Q}$.

\section{Conclusions}

We have conducted momentum-resolved measurements of the phonons in parent and electron-doped \SrIrO{}. In both compounds, our IXS spectra are well reproduced by a nonmagnetic DFT calculation in the LDA, despite the destruction of the spin-orbit assisted Mott-insulating and long-range ordered antiferromagnetic ground state in the latter. In the parent compound, there is no apparent change in the linewidths of the modes on passing through the N\'eel temperature, while the slight changes in frequencies and intensities are fully accounted for by anharmonic interactions and the calculated Debye-Waller factors respectively. In the doped compound, the dispersions of all the measured modes are identical within experimental resolution at both \SI{9}{\kelvin} and \SI{250}{\kelvin}, again with no softening or linewidth changes apparent, nor any peaks in the quasielastic intensity.

Knowledge of the lattice dynamics and their momentum dependence is fundamental to the understanding of the structural, electronic, and magnetic behavior of \SrLaIrO{}. Therefore, our measurements will be important in guiding future theoretical and experimental investigations into the coupled degrees of freedom of this topical material.

\begin{acknowledgements}

C.D.D.~ thanks Atsushi Togo for assistance with the \textsc{phonopy} calculations. C.D.D.~was supported by the Engineering and Physical Sciences Research Council (EPSRC) Centre for Doctoral Training in the Advanced Characterisation of Materials under Grant No.~EP/L015277/1. The IXS measurements were supported by the U.S.~Department of Energy (DOE), Office of Basic Energy Sciences, Early Career Award Program under Award No.~1047478. The DFT calculations were carried out using high performance computing resources at Moscow State University \cite{Sadovnichy2013}. Work at Brookhaven National Laboratory was supported by the U.S.~DOE, Office of Science, Office of Basic Energy Sciences, under Contract No.~DE-SC00112704. Work at UCL was supported by the EPSRC under Grants No.~EP/N027671/1 and EP/N034872/1. Work at Ural Federal University was supported by the Russian Science Foundation under Grant No.~18-12-00185. G.C.~was supported by the U.S.~National Science Foundation under Grant No.~DMR-1712101. The IXS experiments were performed at beamline BL43LXU at the SPring-8 synchrotron with the approval of RIKEN under Proposal No.~20180059.

\end{acknowledgements}

\appendix*

\section{Density Functional Theory Calculations}

\begin{figure*}
	\centering
	\includegraphics[width=\linewidth]{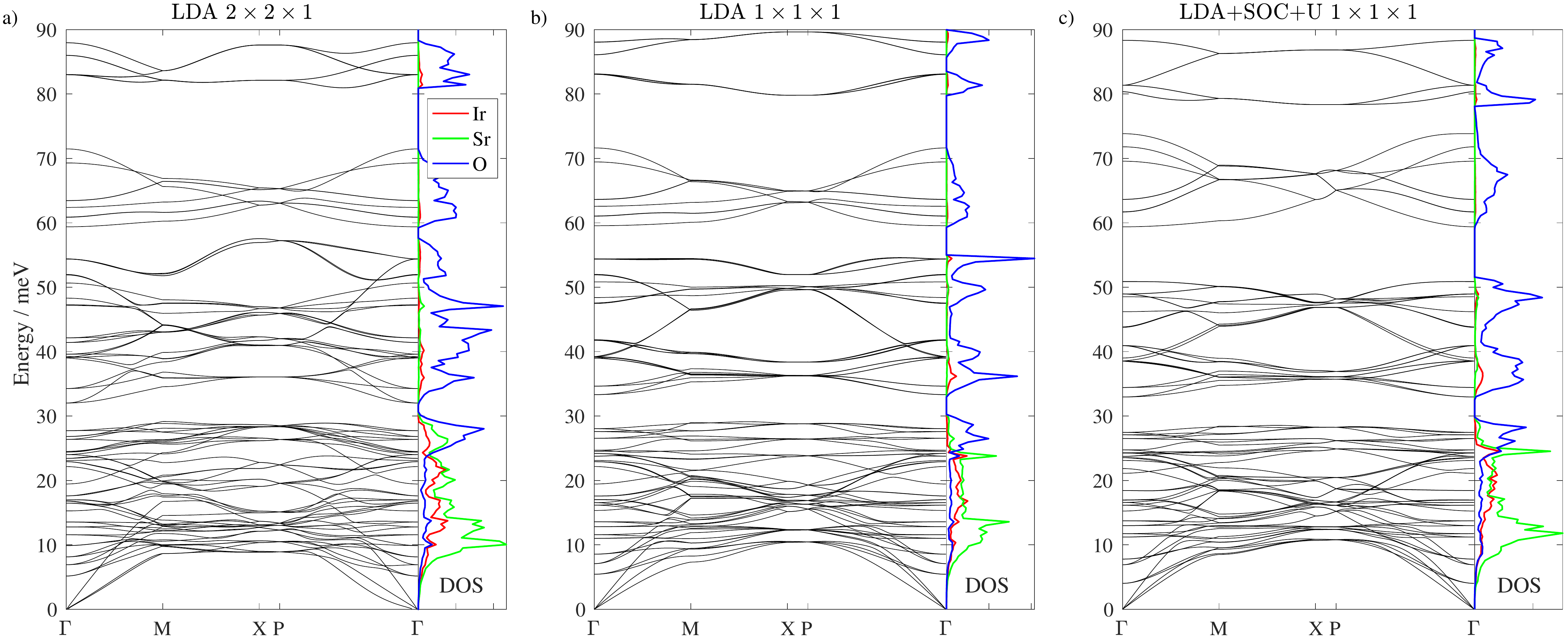}
	\caption{Phonon band structures (black lines) and projected phonon DOS for Ir (red), Sr (green), and O (blue), calculated (a) in the LDA on a $2 \times 2 \times 1$ supercell, (b) in the LDA on a $1 \times 1 \times 1$ supercell, and (c) with LDA+SOC+U on a $1 \times 1 \times 1$ supercell. The addition of SOC+U results in small changes in the phonon dispersions, most prominently in the O modes above \SI{30}{\milli\electronvolt}, while the increasing the cell size leads to far more significant changes.}
	\label{fig:DFT_comp}
\end{figure*}

The DFT calculation in the LDA discussed above did not take into account the effects SOC, U, or the magnetic structure, and therefore does not predict an electronic structure in agreement with the known spin-orbit Mott-insulating state of parent \SrIrO{} (the metallic ground state that it predicts is in fact a better description of the doped compound). We repeated this calculation including SOC+U, using $U =$ \SI{3.05}{\electronvolt} and $J =$ \SI{0.48}{\electronvolt} to reproduce the measured charge gap \cite{Solovyev2015}, and the noncollinear antiferromagnetic structure given by Ye \textit{et al.} \cite{Ye2013}. Due to the additional memory requirements of this calculation, however, the supercell had to be reduced to $1 \times 1 \times 1$.

\begin{figure}
	\centering
	\includegraphics[width=\linewidth]{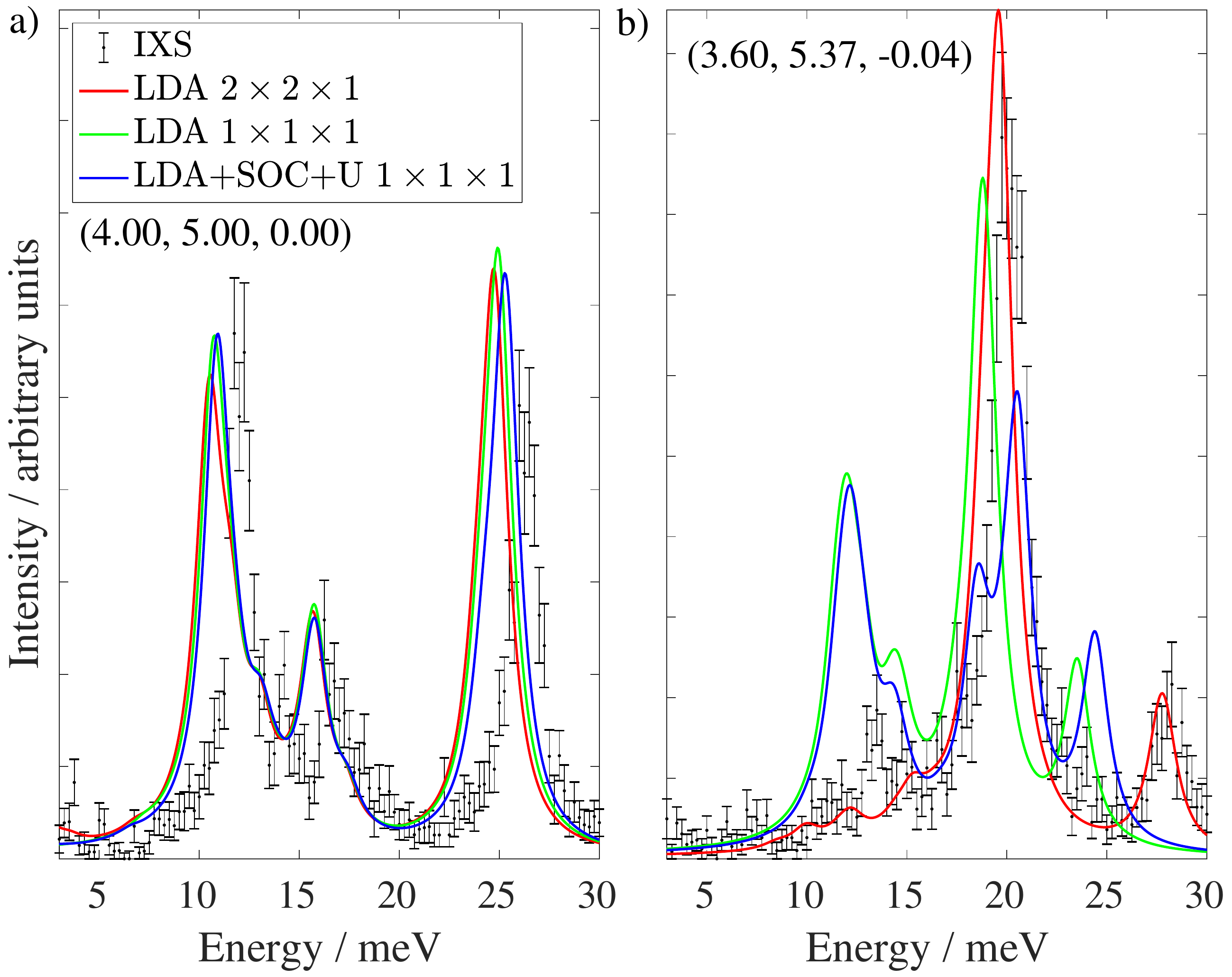}
	\caption{Comparison of IXS spectra (black points) to $S(\mathbf{Q}, \omega)$ calculated with LDA on a $2 \times 2 \times 1$ supercell (red), LDA on a $1 \times 1 \times 1$ supercell (green), and LDA+SOC+U on a $1 \times 1 \times 1$ supercell (blue) at momentum transfers of (a) $(4, 5, 0)$ and (b) $(3.60, 5.37, -0.04)$. At the zone center the calculations differ by only a small shift in the mode energies, while further out in the Brillouin zone the $1 \times 1 \times 1$ supercell calculations predict intense modes that are not present in the experimental data.}
	\label{fig:DFT_spectra_comp}
\end{figure}

Figure \ref{fig:DFT_comp} shows the phonon band structures and projected DOS for these two calculations, as well as for an LDA calculation on a $1 \times 1 \times 1$ supercell for comparison. Comparing the calculations on the minimal unit cell in Fig.~\ref{fig:DFT_comp}(b) and (c), it can be seen that the addition of SOC+U mostly affects the O modes above \SI{30}{\milli\electronvolt}, whose IXS intensities were too low to be measured in this work. Comparing these to the LDA calculation on the larger supercell in Fig.~\ref{fig:DFT_comp}(a), by contrast, shows more significant changes across all of the modes.

We also calculated the dynamic structure factor for these three different cases in order to compare with our IXS measurements. Figure \ref{fig:DFT_spectra_comp} shows this comparison at two representative momentum transfers. At the zone center the supercell size should be of minimal importance, and Fig.~\ref{fig:DFT_spectra_comp}(a) shows that the calculations differ by only a small shift in the mode energies ($\sim$\SI{0.5}{\milli\electronvolt}) and intensities. Further out into the Brillouin zone, however, the supercell size becomes critical, with the calculations on the minimal supercell showing prominent modes that are not present in the IXS spectra [Fig.~\ref{fig:DFT_spectra_comp}(b)].

These comparisons highlight the inadequacy of the minimal supercell in simulating the lattice dynamics of \SrIrO{}. In $5d$ oxides with large unit cells such as these, therefore, where LDA+SOC+U calculations involving noncollinear magnetic structures are prohibitively computationally demanding for larger supercells, a DFT perturbation approach may be more suitable.

\bibliography{Sr214_phonon_PRB}

\end{document}


\title{Supplemental Material: Momentum-resolved lattice dynamics of parent and electron-doped \SrIrO{}} 

\author{C.~D.~Dashwood}
\email{cameron.dashwood.17@ucl.ac.uk}
\affiliation{London Centre for Nanotechnology and Department of Physics and Astronomy, University College London, Gower Street, London WC1E 6BT, UK}

\author{H.~Miao}
\affiliation{Department of Condensed Matter Physics and Materials Science, Brookhaven National Laboratory, Upton, New York 11973, USA}

\author{J.~G.~Vale}
\affiliation{London Centre for Nanotechnology and Department of Physics and Astronomy, University College London, Gower Street, London WC1E 6BT, UK}

\author{D.~Ishikawa}
\affiliation{Materials Dynamics Laboratory, RIKEN SPring-8 Center, RIKEN, Sayo Hyogo 697-5148, Japan}

\author{D.~A.~Prishchenko}
\affiliation{Department of Theoretical Physics and Applied Mathematics, Ural Federal University, 19 Mira Street, Ekaterinburg 620002, Russia}

\author{V.~V.~Mazurenko}
\affiliation{Department of Theoretical Physics and Applied Mathematics, Ural Federal University, 19 Mira Street, Ekaterinburg 620002, Russia}

\author{V.~G.~Mazurenko}
\affiliation{Department of Theoretical Physics and Applied Mathematics, Ural Federal University, 19 Mira Street, Ekaterinburg 620002, Russia}

\author{R.~S.~Perry}
\affiliation{London Centre for Nanotechnology and Department of Physics and Astronomy, University College London, Gower Street, London WC1E 6BT, UK}

\author{G.~Cao}
\affiliation{Department of Physics, University of Colorado at Boulder, Boulder, Colorado 80309, USA}

\author{A.~de la Torre}
\affiliation{Institute for Quantum Information and Matter and Department of Physics, California Institute of Technology, Pasadena, California 91125, USA}
\affiliation{Department of Quantum Matter Physics, University of Geneva, 24 Quai Ernest-Ansermet, 1211 Geneva 4, Switzerland}

\author{F.~Baumberger}
\affiliation{Department of Quantum Matter Physics, University of Geneva, 24 Quai Ernest-Ansermet, 1211 Geneva 4, Switzerland}

\author{A.~Q.~R.~Baron}
\affiliation{Materials Dynamics Laboratory, RIKEN SPring-8 Center, RIKEN, Sayo Hyogo 697-5148, Japan}

\author{D.~F.~McMorrow}
\address{London Centre for Nanotechnology and Department of Physics and Astronomy, University College London, Gower Street, London WC1E 6BT, UK}

\author{M.~P.~M.~Dean}
\affiliation{Department of Condensed Matter Physics and Materials Science, Brookhaven National Laboratory, Upton, New York 11973, USA}

\date{\today}

\maketitle

\beginsupplement

\section{Dynamic structure factor calculations}

The intensity measured in non-resonant inelastic x-ray scattering (IXS) at total momentum transfer $\mathbf{Q}$ and energy loss $\omega$ is directly proportional to the dynamic structure factor $S(\mathbf{Q}, \omega)$, with the constant of proportionality depending on the momenta and polarisation of the incoming/outgoing photons \cite{Baron2009}. Away from Bragg reflections, $S(\mathbf{Q}, \omega)$ is dominated by the one-phonon term, which in the Born approximation reads
\begin{equation}
	\displaystyle S(\mathbf{Q}, \omega) = \sum_{q, j} \left|F(\mathbf{Q}, \mathbf{q}, j)\right|^2 \langle n_{qj} + 1\rangle \delta(\omega - \omega_{qj}) \delta(\mathbf{Q} - \mathbf{q})
\label{equ:S}
\end{equation}
where $\mathbf{q}$ sums over the reduced momenta in the first Brillouin zone, and $j$ sums over the $3n$ modes at $\mathbf{q}$ ($n$ being the number of atoms in the primitive unit cell) with frequencies $\omega_{qj}$ and Bose factors $\langle n_{qj} + 1\rangle = \left(1 - e^{-\omega_{qj} / (k_bT)}\right)^{-1}$. $F$ is given by
\begin{equation}
	\displaystyle F(\mathbf{Q}, \mathbf{q}, j) = \sum_d \sqrt{\frac{f_d(\mathbf{Q})}{2 m_d \omega_{qj}}} e^{-W_d(\mathbf{Q})} e^{-\i\mathbf{Q} \cdot \mathbf{r}_d} \mathbf{Q} \cdot \mathbf{e}_{qjd}
\label{equ:F}
\end{equation}
where $d$ sums over the $n$ atoms in the primitive unit cell with masses $m_d$ located at $\mathbf{r}_d$, $e^{-W_d(\mathbf{Q})}$ is the Debye-Waller factor, $\mathbf{e}_{qjd}$ the complex phonon polarisation, and $f_d(\mathbf{Q})$ the x-ray form factor.

Equation \ref{equ:S} was evaluated using the \textsc{phonopy} package \cite{Togo2015}. In order to be directly compared with our IXS data, however, a finite linewidth $\gamma_{qj}$ (due mostly to instrument resolution) must be introduced by replacing the frequency delta function with a damped harmonic oscillator lineshape
\begin{equation}
	\displaystyle \delta(\omega - \omega_{qj}) \rightarrow \chi_j''(\mathbf{q}, \omega) = \frac{4 \omega \omega_{qj} \gamma_{qj}}{\pi \left[\left(\omega^2 - \omega_{qj}^2 - \gamma_{qj}^2\right)^2 + 4 \omega^2 \gamma_{qj}^2\right]}.
\label{equ:DHO}
\end{equation}
We use $\gamma_{qj} = \gamma =$ \SI{0.8}{\milli\electronvolt} for all modes at all momenta.

\clearpage

\section{Space group comparison}

\begin{figure}[b]
	\centering
	\includegraphics[width=\linewidth]{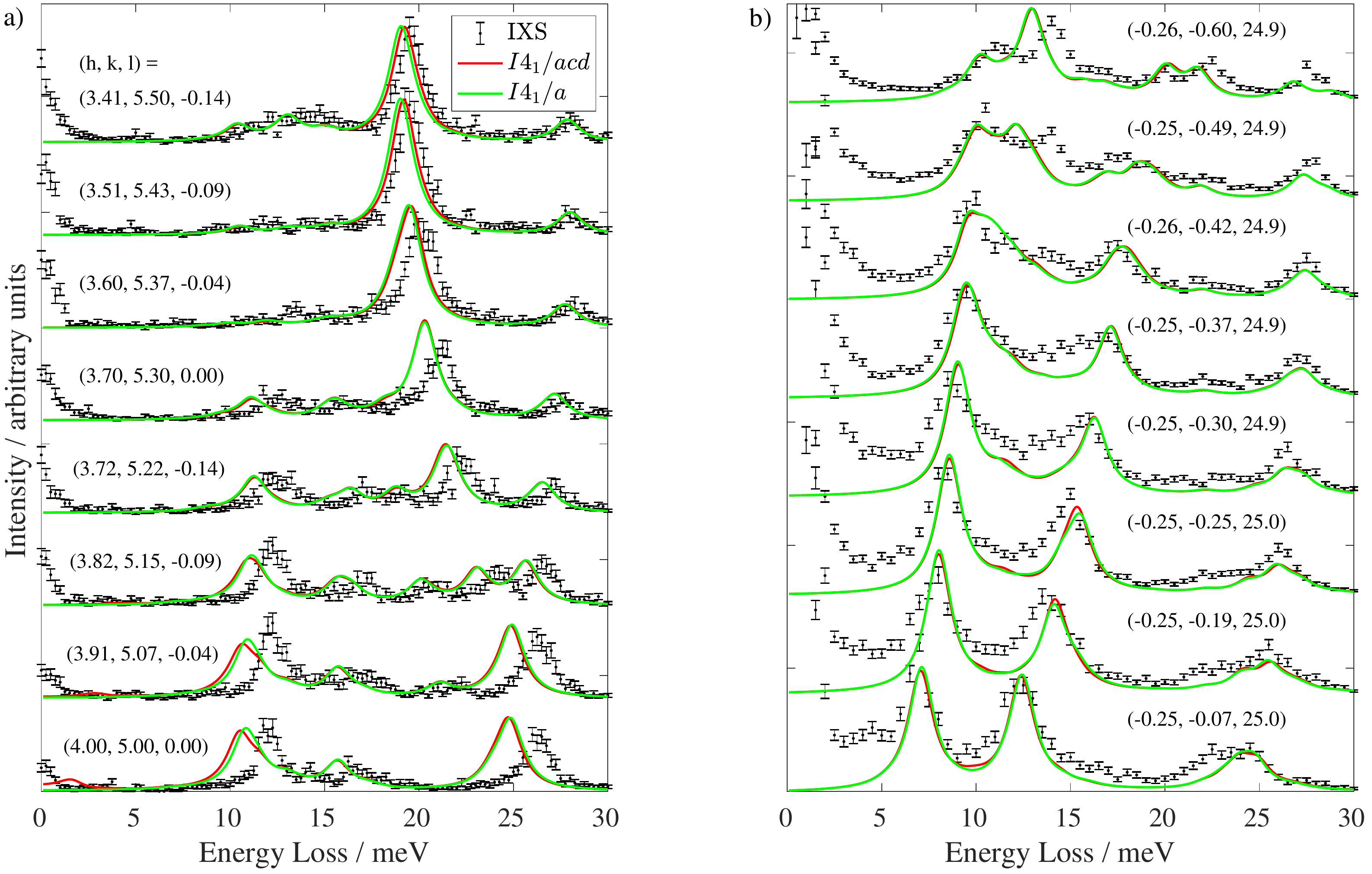}
	\caption{Comparison of the DFT calculations for the $I4_1/acd$ (red lines) and $I4_1/a$ (green lines) space groups with the IXS data (black points) for a) the parent compound at \SI{100}{\kelvin} and b) the doped compound at \SI{9}{\kelvin}. All calculations were performed in the LDA approximation on a $2\times2\times1$ supercell.}
	\label{fig:space_group_comp}
\end{figure}

Neutron diffraction \cite{Ye2013, Dhital2013} and nonlinear optical experiments \cite{Torchinsky2015} have shown the true space group of \SrIrO{} to be $I4_1/a$. The difference from $I4_1/acd$ is very subtle, however, consisting of a relative difference of $\sim10^{-3}$ in the tetragonal distortions of the oxygen octahedra around the iridium ions on each of the two sub-lattices. Although it has been shown that this structural distortion is important for the locking of the magnetic moment to the octahedral rotation \cite{Torchinsky2015}, we believe that it should be of minimal importance for the calculated phonon dispersions. To show this explicitly, we have repeated the DFT calculations using the $I4_1/a$ structure from Ye \textit{et al.} \cite{Ye2015}. Figure \ref{fig:space_group_comp} shows a comparison between the $I4_1/acd$ and $I4_1/a$ calculations. It can be seen that there is very little difference between the two, with shifts in the phonon energies well below the resolution of our IXS measurements.

\clearpage

\section{Comparison with optical measurements}

Here we compare our LDA calculation on the $2 \times 2 \times 1$ supercell with previously published Raman and infra-red (IR) spectroscopy studies. The primitive unit cell of \SrIrO{} contains 28 atoms, giving 84 phonon modes. A symmetry analysis of the $I4_1/acd$ space group with Ir atoms at Wyckoff site $8a$, Sr at $16d$ and O at $16d$ (apical) and $16f$ (basal) shows that the Raman active modes have irreducible representations $3 A_{1g} + 5 B_{1g} + 4 B_{2g} + 13 E_g$ and the IR modes have irreducible representations $5 A_{2u} + 13 E_u$, leaving $4 A_{1u} + 4 A_{2g} + 4 B_{1u} + 3 B_{2u}$. Table \ref{tab:Raman_IR_comp} shows that our calculated mode energies compare well to the Raman and IR studies.

\begin{longtable}{ccc|c|cc|cc}
		\hline \hline
		\multicolumn{3}{c|}{LDA $2 \times 2 \times 2$} & Ref.~\onlinecite{Samanta2018} & \multicolumn{2}{c|}{Ref.~\onlinecite{Gretarsson2017} (Raman)} & \multicolumn{2}{c}{Ref.~\onlinecite{Propper2016} (IR)} \\
		Band & Assignment & Calculated & Calculated & Calculated & Observed & Calculated & Observed \\
		\hline
		1, 2 & $E_{u}$ & Acoustic & Acoustic & - & - & - & - \\
		3 & $A_{2u}$ & Acoustic & Acoustic & - & - & - & - \\
		4, 5 & $E_{u}$ & 42 & Imaginary & - & - & 30 & - \\
		6, 7 & $E_{g}$ & 56 & Imaginary & - & - & - & - \\
		8, 9 & $E_{g}$ & 66 & 53 & - & - & - & - \\
		10, 11 & $E_{u}$ & 80 & 84 & - & - & 81 & 103 \\
		12, 13 & $E_{g}$ & 93 & 91 & - & - & - & - \\
		14, 15 & $E_{u}$ & 103 & 101 & - & - & 92 & 115 \\
		16, 17 & $E_{g}$ & 109 & 108 & - & - & - & - \\
		18 & $B_{1g}$ & 123 & 114 & 110 & - & - & - \\
		19, 20 & $E_{g}$ & 133 & 132 & - & - & - & - \\
		21, 22 & $E_{u}$ & 135 & 120 & - & - & 122 & 138 \\
		23 & $B_{2g}$ & 137 & 134 & 135 & - & - & - \\
		24 & $A_{1u}$ & 137 & - & - & - & - & - \\
		25 & $A_{2g}$ & 142 & - & - & - & - & - \\
		26 & $B_{1u}$ & 142 & - & - & - & - & - \\
		27 & $B_{2u}$ & 157 & - & - & - & - & - \\
		28 & $B_{1g}$ & 179 & 167 & 168 & - & - & - \\
		29 & $A_{1u}$ & 185 & - & - & - & - & - \\
		30 & $B_{2g}$ & 185 & 173 & - & - & - & - \\
		31 & $A_{2u}$ & 186 & 172 & - & - & 172 & 192 \\
		32, 33 & $E_{u}$ & 189 & 174 & - & - & 178 & - \\
		34, 35 & $E_{g}$ & 194 & 186 & 171 & 191 & - & - \\
		36, 37 & $E_{u}$ & 197 & 194 & - & - & 186 & - \\
		38 & $A_{1g}$ & 198 & 181 & 188 & 188 & - & - \\
		39, 40 & $E_{g}$ & 213 & 203 & - & - & - & - \\
		41, 42 & $E_{u}$ & 216 & 204 & - & - &212 & 214 \\
		43, 44 & $E_{g}$ & 224 & 205 & - & - & - & - \\
		45, 46 & $E_{u}$ & 258 & 257 & - & - & 251 & 270 \\
		47, 48 & $E_{g}$ & 276 & 266 & - & - & - & - \\
		49 & $B_{2u}$ & 315 & - & - & - & - & - \\
		50, 51 & $E_{g}$ & 315 & 291 & - & - & - & - \\
		52 & $A_{1g}$ & 316 & 260 & 326 & 278 & - & - \\
		53, 54 & $E_{u}$ & 319 & 293 & - & - & 298 & 283 \\
		55, 56 & $E_{g}$ & 334 & 314 & - & - & - & - \\
		57, 58 & $E_{u}$ & 340 & 314 & - & - & 323 & 324 \\
		59 & $A_{2g}$ & 380 & - & - & - & - & - \\
		60 & $B_{1u}$ & 381 & - & - & - & - & - \\
		61 & $A_{2u}$ & 390 & 340 & - & - & 374 & 373 \\
		62 & $B_{1g}$ & 409 & 359 & 391 & - & - & - \\
		63 & $A_{1u}$ & 419 & - & - & - & - & - \\
		64 & $B_{2g}$ & 419 & 371 & 410 & 395 & - & - \\
		65, 66 & $E_{u}$ & 439 & 381 & - & - & 406 & 367 \\
		67, 68 & $E_{g}$ & 439 & 380 & - & - & - & - \\
		69 & $A_{2u}$ & 479 & 502 & - & - & 443 & 515 \\
		70 & $B_{2g}$ & 491 & 513 & 457 & 495 & - & - \\
		71 & $A_{1u}$ & 491 & - & - & - & - & - \\
		72 & $B_{2u}$ & 503 & - & - & - & - & - \\
		73 & $B_{1u}$ & 512 & - & - & - & - & - \\
		74 & $A_{2g}$ & 512 & - & - & - & - & - \\
		75 & $A_{1g}$ & 559 & 588 & 532 & 562 & - & - \\
		76 & $B_{1g}$ & 576 & 596 & 546 & - & - & - \\
		77, 78 & $E_{u}$ & 669 & 751 & - & - & 645 & 664 \\
		79, 80 & $E_{g}$ & 670 & 751 & - & - & - & - \\
		81 & $B_{1g}$ & 694 & 808 & 660 & 693 & - & - \\
		82 & $A_{2u}$ & 694 & 808 & - & - & 660 & - \\
		83 & $B_{1u}$ & 709 & - & - & - & - & - \\
		84 & $A_{2g}$ & 709 & - & - & - & - & - \\
		\hline \hline
		\caption{Comparison of the modes calculated with LDA on a $2 \times 2 \times 1$ supercell to those from previously published Raman and IR studies on \SrIrO{}. All frequencies are in \SI{}{\per\centi\meter}. The imaginary frequencies in Ref.~\onlinecite{Samanta2018} arise from their use of the unrelaxed experimental crystal structure.}
		\label{tab:Raman_IR_comp}
\end{longtable}

\clearpage

\section{Supplemental data for the parent and doped samples}

\begin{figure}
	\centering
	\includegraphics[width=\linewidth]{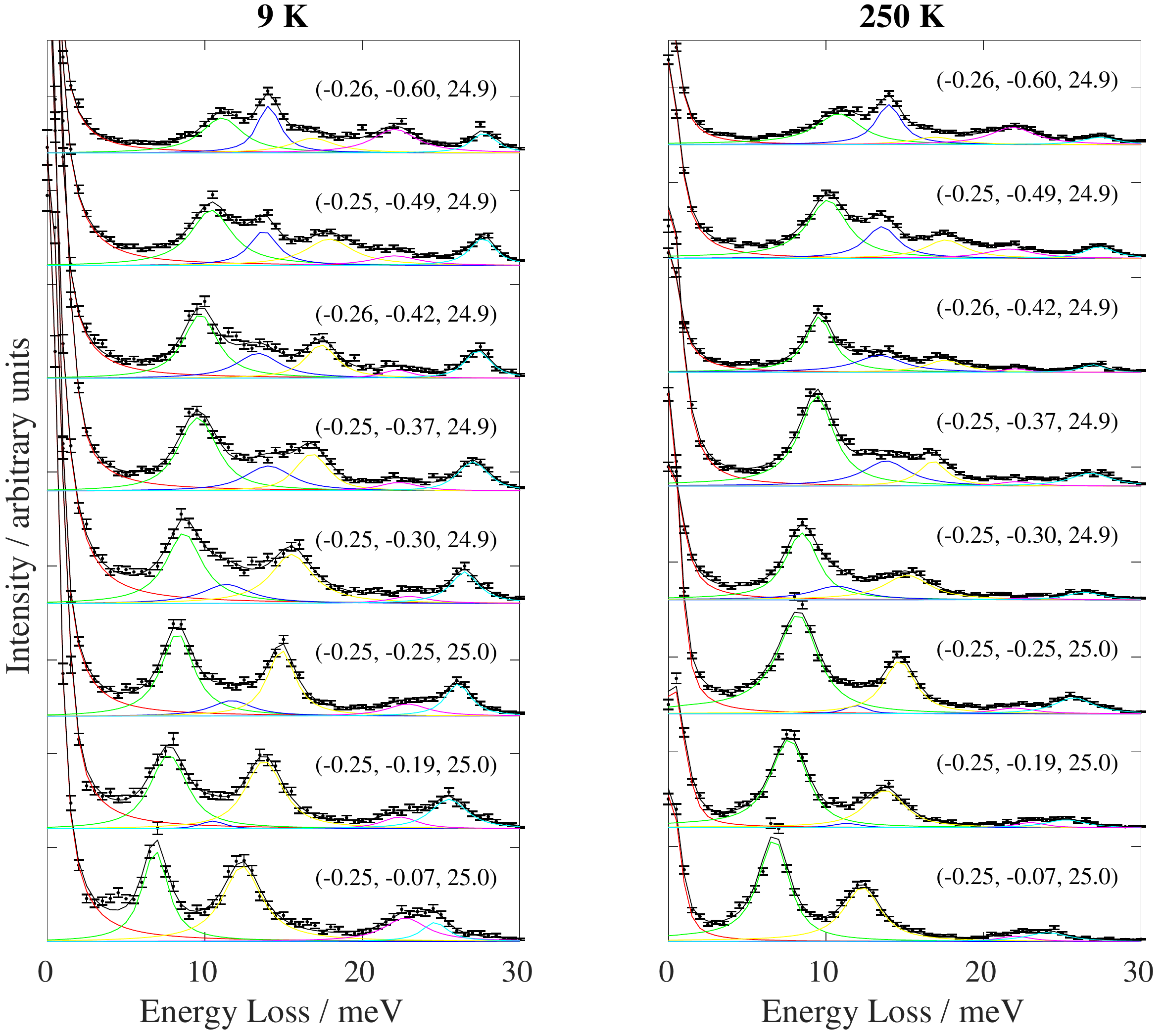}
	\caption{Detail of the fits (black line) of the doped spectra (black points). As described in the main text, the elastic peak is fitted with a pseudo-Voigt resolution function (red), while each of the phonon modes is fitted with a damped harmonic oscillator lineshape weighted by the Bose factor and convoluted with the resolution function (green, blue, yellow, magenta and cyan respectively for each mode).}
	\label{fig:doped_fit}
\end{figure}

Here we present IXS spectra and LDA calculations on the $2 \times 2 \times 1$ supercell for the $\mathbf{Q}$ points indicated in Fig. \ref{fig:DFT} but not shown in the main text. The conclusions drawn in the main text (namely the good agreement between the experimental spectra and non-magnetic DFT calculations, and the lack of any visible anomalous frequency or linewidth changes through $T_N$) apply equally well to these data. Fig. \ref{fig:doped_fit} shows details of the fits of the doped spectra at \SI{9}{\kelvin} and \SI{250}{\kelvin}.

\begin{sidewaysfigure}
	\centering
	\includegraphics[width=\linewidth]{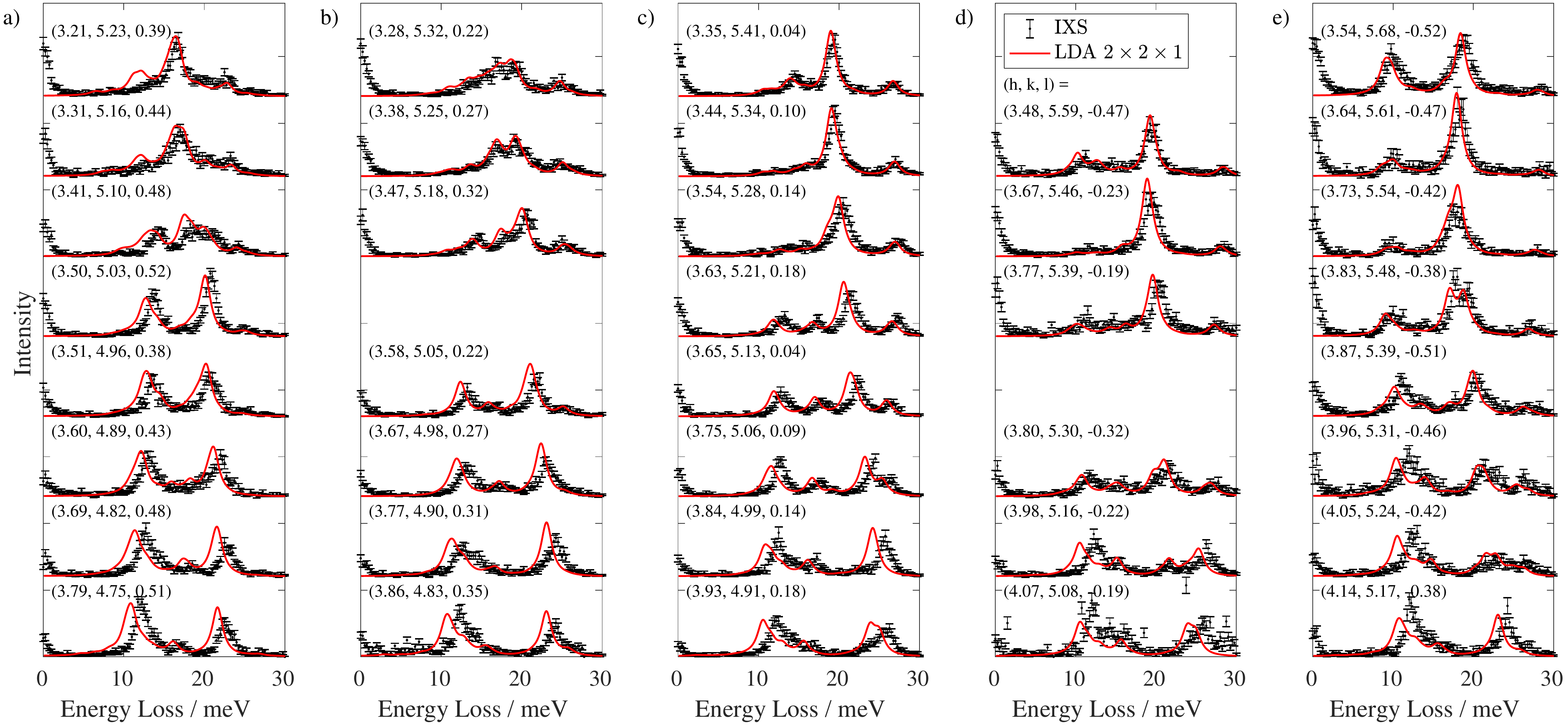}
	\caption{IXS spectra in the parent compound at 100 K for each vertical column of analysers not shown in the main text (black points) compared to the dynamic structure factor calculated with LDA on a $2 \times 2 \times 1$ supercell (solid red lines). The spectra in each plot are offset vertically for clarity. The missing spectra are due to detector malfunctions.}
	\label{fig:parent_spectra_full}
\end{sidewaysfigure}

\begin{sidewaysfigure}
	\centering
	\includegraphics[width=\linewidth]{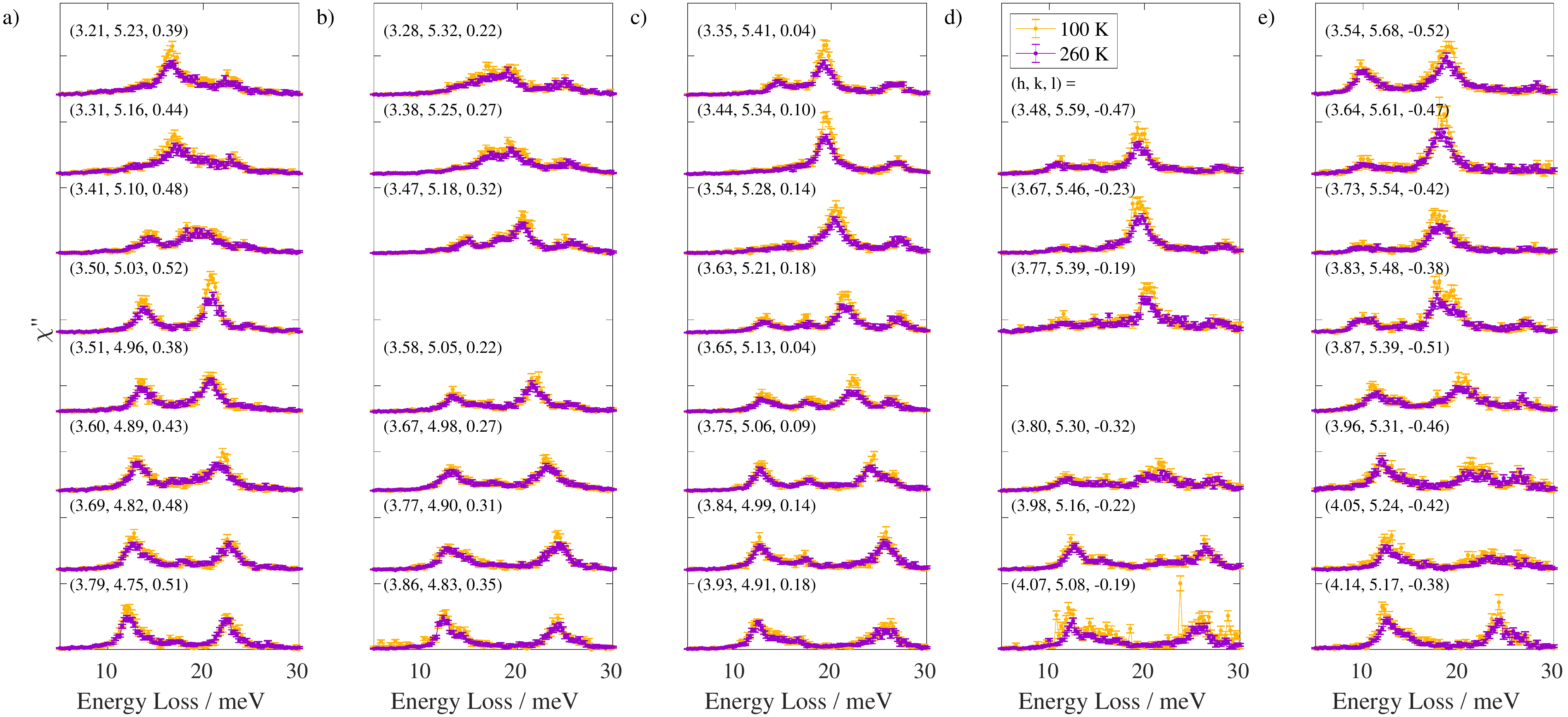}
	\caption{Bose-factor corrected IXS spectra in the parent compound at 100 K (orange) and 260 K (purple) for for each vertical column of analysers not shown in the main text. The spectra in each plot are offset vertically for clarity. The missing spectra are due to detector malfunctions.}
	\label{fig:parent_temp_comp_full}
\end{sidewaysfigure}

\begin{sidewaysfigure}
	\centering
	\includegraphics[width=\linewidth]{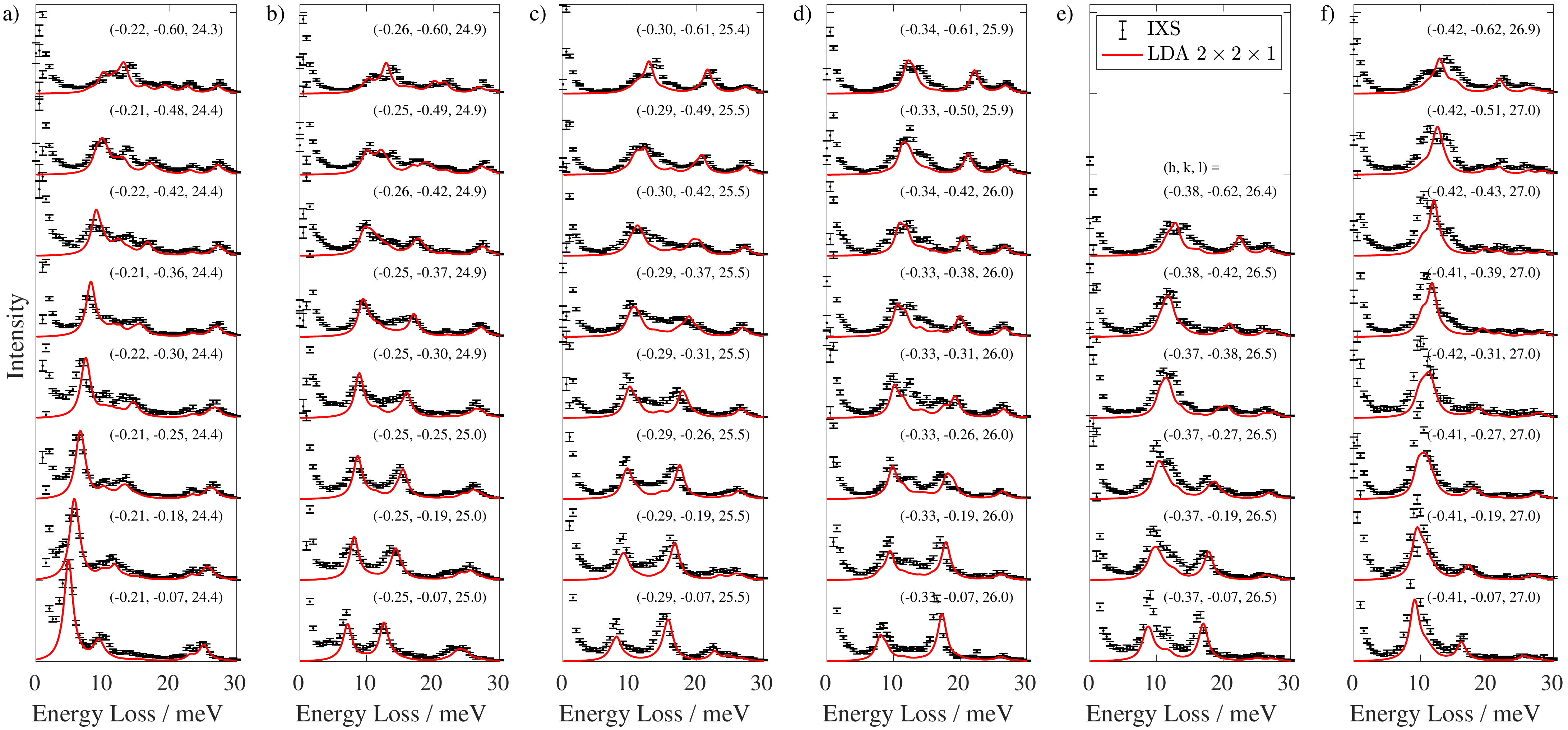}
	\caption{IXS spectra in the doped compound at 9 K for each vertical column of analysers not shown in the main text (black points) compared to the dynamic structure factor calculated with LDA on a $2 \times 2 \times 1$ supercell (solid red lines). The spectra in each plot are offset vertically for clarity. The missing spectra are due to detector malfunctions.}
	\label{fig:doped_spectra}
\end{sidewaysfigure}

\begin{sidewaysfigure}
	\centering
	\includegraphics[width=\linewidth]{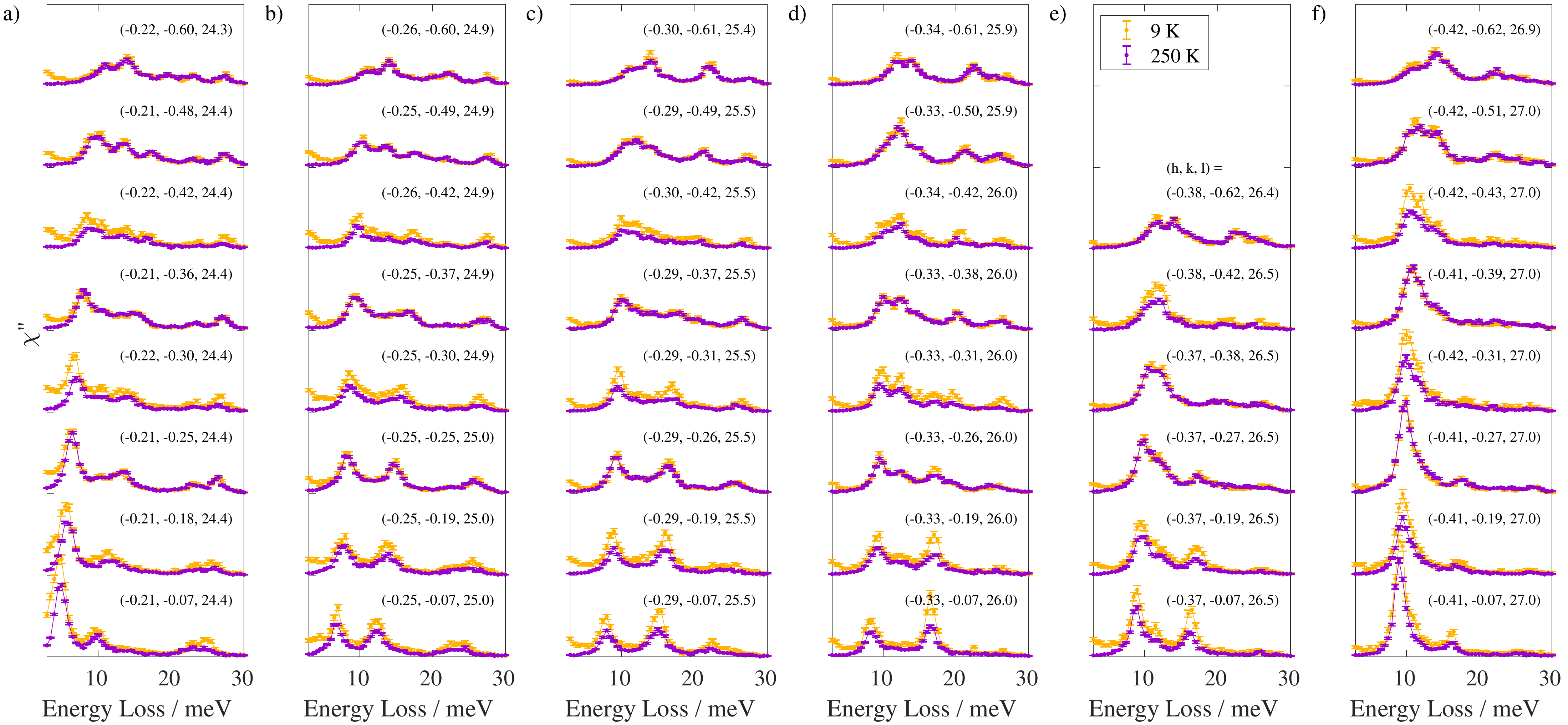}
	\caption{Bose-factor corrected IXS spectra in the doped compound at 9 K (orange) and 250 K (purple) for for each vertical column of analysers not shown in the main text. The spectra in each plot are offset vertically for clarity. The missing spectra are due to detector malfunctions.}
	\label{fig:doped_temp_comp}
\end{sidewaysfigure}

\bibliography{../Sr214_phonon_PRB}